\documentstyle[12pt,a4]{article}
\newcommand{\be}{\begin{eqnarray*}}
\newcommand{\bet}{\begin{center} \begin{tabular}}
\newcommand{\bit}{\begin{itemize}}
\newcommand{\ee}{\end{eqnarray*}}
\newcommand{\eit}{\end{itemize}}

\newcommand{\ent}{\end{tabular} \end{center}}

\newcommand{\half}{\frac{1}{2}}

\newcommand{\eennww}{e^- e^- \rightarrow W^- W^- \nu_e \nu_e}
\newcommand{\eewzen}{e^- e^- \rightarrow W^- Z^0 e^- \nu_e}
\newcommand{\eezzee}{e^- e^- \rightarrow Z^0 Z^0 e^- e^-}
\newcommand{\WW}{W^-W^-\nu_e\nu_e}
\newcommand{\WZ}{W^-Z^0e^-\nu_e}
\newcommand{\ZZ}{Z^0Z^0e^-e^-}
\newcommand{\bb}{\begin{thebibliography}}
\newcommand{\eb}{\end{thebibliography}}

\newcommand{\bi}[1]{\bibitem{#1}}

\newcommand{\beq}{\begin{equation}}
\newcommand{\eeq}{\end{equation}}
\newcommand{\bea}{\begin{eqnarray}}
\newcommand{\eea}{\end{eqnarray}}

\newcommand{\nn}{\nonumber}
\newcommand{\ra}{\rightarrow}

\newcommand{\sm}{standard model }

\newcommand{\prd}{Phys.\ Rev.\ D }

\newcommand{\np}{Nucl.\ Phys.\ B }
\newcommand{\zp}{Z.\ Phys.\ C }

\newcommand{\plb}{Phys.\ Lett.\ B }

\begin{document}
\thispagestyle{empty}
\setcounter{page}{0}

\begin{flushright}
MPI-PhT/94-27\\
LMU-06/94\\
May 1994
\end{flushright}

\vspace*{15mm}

\begin{center}

{\Large \bf Standard Model Predictions for Weak Boson Pair Production
in $e^-e^-$ Scattering \par}

\vspace{16mm}

{\large
F.~Cuypers$^{\rm a}$,
K.~Ko\l odziej$^{\rm b,}$\footnote{
        Work supported by the German Federal Ministry for Research
        and Technology under contract No.~05 6MU 93P}$^,$\footnote{
        On leave from the Institute of Physics, University of Silesia,
        ul.~Uniwersytecka 4, PL--40007 Katowice, Poland}
and R.~R\"uckl$^{\rm a,b,1}$} \\

\bigskip
{\em $^{\rm a}$Max-Planck-Institut f\"ur Physik, Werner-Heisenberg-Institut,\\
               F\"ohringer Ring 6, D--80805 M\"unchen, FRG}\\
\medskip
{\em $^{\rm b}$Sektion Physik der Universit\"at M\"unchen, \\
               Theresienstr.~37, D--80333 M\"unchen, FRG}\\

\end{center}

\vspace*{16mm}

\centerline{\bf Abstract}
\noindent
\normalsize
We study polarized $e^-e^-$ scattering into $\WW$, $\WZ$ and
$\ZZ$ final states within the framework of the standard model
of electroweak interactions.
These mechanisms for producing pairs of weak gauge bosons are potential
backgrounds to new reactions
beyond the realm of the standard model.
At a centre-of-mass energy of 500 GeV the total cross sections are
calculated to be 2.5, 9.4 and 1.0 fb, respectively, for unpolarized
beams. The energy behaviour of the cross sections
as well as particularly interesting differential distributions are presented
and the topology of the final states is discussed.
\vfill
\newpage

\section{Introduction}

It is becoming clear that a linear collider
will be one of the most promising tools
for high energy experimentation
in the begin of the next century.
A number of world-wide collaborations
is already developing different designs
aiming at a 0.5 -- 2  TeV machine
capable of delivering 10 fb$^{-1}$ of luminosity per year.
%and possible further upgrades.
The possibility of operating a linear collider in $e^+e^-$,
$e^-e^-$, $e\gamma$ or $\gamma\gamma$ modes \cite{Ginzburg},
with high degrees of polarization \cite{workshop}, is one of the
big advantages of this new facility.

An interesting feature special to $e^-e^-$ reactions
is their particular sensitivity to lepton-number violating processes.
For example,
a resonance in the channel  $e^-e^- \rightarrow \mu^-\mu^-$
would reveal the presence of dileptons \cite{Frampton},
{\em i.e.} doubly charged gauge bosons
which appear naturally in a wide class of gauge extensions of the \sm.
Another very important example is the observation of $W$-pair production in
$e^-e^- \rightarrow W^-W^-$ which
would provide strong evidence
for the existence of Majorana neutrinos \cite{Minkowski}.
Furthermore,
the production of chargino or selectron pairs
via sneutrino or neutralino exchanges in
$e^-e^- \rightarrow \tilde{\chi}^- \tilde{\chi}^-$
or
$e^-e^- \rightarrow \tilde{e}^- \tilde{e}^-$
may serve as one of the most powerful probes
for supersymmetry \cite{KL,COR1}.

But also processes which do not violate lepton number conservation
can be of interest in $e^-e^-$ collisions.
Indeed,
the reaction $e^-e^- \rightarrow e^-W^-\nu_e$
provides informations on possible anomalous boson couplings
which are complementary to those  obtainable
in $e^+e^-$, $e\gamma$ and $\gamma\gamma$ scattering \cite{wo1}.
Furthermore,
the existence of an extra $Z'$ gauge boson
can be better probed in M\o ller scattering
than with $e^+e^-$ collisions \cite{wo2}.

In order to evaluate this new physics potential
it is necessary to understand the standard model processes very accurately.
We have begun to study this problem in a recent publication \cite{CKKR},
where we have reported results on the total cross section
and important differential distributions for the reaction $\eennww$.
Here we extend this study to $\WZ$ and $\ZZ$ final states
and provide a comprehensive account of our results on
pair production of weak gauge bosons in $e^-e^-$ scattering.
Obviously these reactions are potential
sources of background to the exotic reactions mentioned above,
if the $Z^0$ bosons decay invisibly
or electrons are lost along the beam-pipe.
In addition, gauge boson pair production may also become interesting in its own
right
in the TeV energy regime,
because of subtle $SU(2)_L \times U(1)_Y$ gauge cancellations
which depend on the quartic gauge coupling of $W$ bosons.
Preliminary results of our study have been reported
at a recent workshop \cite{CKR}.

The paper is organized as follows.
In Section~2 we describe
some technicalities of the calculation of the helicity matrix
elements and the phase space integration.
The total cross sections and events rates in the energy range from
500 GeV to 2 TeV are given in Section~3.
In Section~4 we present
some important differential distributions and discuss the general
topology of the final states.
We also indicate kinematical cuts
which would eliminate these \sm\ processes
as backgrounds to `new' physics.
A brief summary and conclusions are given in Section~5.

\section{Helicity Amplitudes and Phase Space Integration}

In Figs~\ref{f1} to \ref{f3} we display the different topologies of the lowest
order Feynman diagrams for the processes
\bea
\label{eq:ee}
e^-(p_1)+e^-(p_2) \ra  \left\{ \begin{array}{cr}
                 W^-(k_1)+W^-(k_2)+\nu_e(q_1)+\nu_e(q_2)~,&
                 \qquad\qquad\qquad\qquad (1)\\[2mm]
                 W^-(k_1)+Z^0(k_2)+e^-(q_1)+\nu_e(q_2)~,&
                 \qquad\qquad\qquad\qquad (2)\\[2mm]
                 Z^0(k_1)+Z^0(k_2)+e^-(q_1)+e^-(q_2)~,&
                 \qquad\qquad\qquad\qquad (3)\nn
                 \end{array}
                 \right.
\eea
where the particle momenta are indicated in parenthesis. The actual
diagrams can be easily obtained from the generic ones given in the
figures by an appropriate particle assignment and by permutations
of the momenta of identical particles.
In the limit of zero electron mass
(when the electron does not couple to the Higgs field),
there are 66 diagrams contributing to the $\WW$ final state.
Since in this case each of the two fermion lines is coupled to at least one
$W^-$ boson, as can be seen in Fig.~\ref{f1}, only the left-left ($LL$)
combination of initial polarizations has a non-zero cross section.
In contrast, the $\WZ$ final states can be produced with $LL$ and left-right
($LR$) beams. Here one has contributions from 88
and 37 Feynman diagrams, respectively. Finally, in the $\ZZ$ channel
all combinations of beam polarizations have finite cross sections.
The number of Feynman diagrams for $LL$, $LR$ and right-right ($RR$)
scattering is 86, 43 and 86, respectively.

The large number of diagrams makes the use of conventional trace
techniques for calculating the matrix element squared rather
uneconomical.
Moreover,
in the high energy regime large gauge cancellations
can cause numerical instabilities
when squares of diagrams and interference terms are added.
It is therefore advisable
to compute directly the helicity amplitudes
associated with definite polarizations of the initial and final states.
In order to warrant to a high degree the correctness of our results,
we have performed two independent calculations,
making use of two different four-body phase space integration routines
and helicity amplitude formalisms.
The first helicity method is the conventional one,
based on the Weyl--van der Waerden formalism.
Although originally developed for massless particles,
it was extended to massive bosons in Ref.~\cite{Kleiss}.
The second method,
described in detail in Ref.~\cite{KZ},
uses the Weyl representation of the Dirac matrices and spinors,
which is particularly suitable for handling massless fermions.
In this representation the spinors have only two non-zero entries
and the Dirac matrices have an off-diagonal form, when represented
by blocks of $2\times 2$ matrices. This allows to reduce the
$4\times 4$ Dirac algebra to the $2\times 2$ Pauli algebra.
Given the initial and finite state polarization, one has to
calculate the relevant elements of a $2\times 2$ matrix for each
Feynman diagram. This matrix is composed of an odd number of Pauli
matrices contracted with some combinations of the particle  momenta
or the gauge boson polarization vectors. For better numerical
performance, the latter are taken to be real.
The results of the two calculations are in perfect numerical agreement.

In addition, for each of the processes (1) to (3) we have checked
internal or external gauge invariance. All calculations have been
performed in a general covariant ($R_{\xi}$) gauge.
For $\eennww$ we verified gauge invariance
with respect to a change of the gauge parameter $\xi$
in the gauge boson propagator.
Numerically we have observed invariance of the matrix element squared
averaged over polarizations up to 12 digits,
when the gauge parameter $\xi$ is varied over the range
$-10^{5} \leq \xi \leq 10^5$.
For $\eewzen$ and $\eezzee$ with the $Z^0$ boson replaced
by a photon, we have checked external gauge invariance. This time
we have observed numerically a drop of the real and imaginary part of
the matrix elements by several orders
of magnitude when the photon polarization vector is replaced by its
momentum.

The electron mass is neglected everywhere,
except in the denominator of the photon propagator whenever it is
needed to regulate collinear singularities.
For instance, in reaction (2), when a photon is coupled to an on-shell
electron line,
the momentum transfer flowing through the photon propagator
can be written as
\setcounter{equation}{3}
\bea
\label{eq:t}
    t = (p_1-q_1)^2 = -\sqrt{s}E' \beta\beta'
        (1 - \cos\theta ) - t^{\rm min}~,
\eea
where the particle momenta are defined in Eq.~(2).
$E'$ is the energy of the final state electron,
$\beta = (1 - {4m_e^2\over s})^{\half}$
and
$\beta' = (1 - {m_e^2\over E'^2})^{\half}$
are the velocities of the initial and final state electrons,
and $\theta$ is the angle of the final state
electron with respect to the beam axis in the centre-of-mass system.
To order $m_e^2$,
the minimal absolute value of the momentum transfer is given by
\bea
\label{eq:tmin}
    t^{\rm min} = m_e^2 {{(\sqrt{s} - 2E')^2} \over {2\sqrt{s}E'}}~.
\eea
To avoid the collinear singularity in the phase space integral of
terms proportional to $1/t$ in the squared matrix element,
one has to keep term (\ref{eq:tmin}) in the
denominator of the photon propagator given by Eq.~(\ref{eq:t}).
This approximate
procedure takes properly into account the leading collinear logarithms,
but neglects constant terms.
The point is that
in addition to the terms proportional to $1/t$ there appear
$m_e^2/t^2$-terms in the
squared matrix elements which give rise to constant contributions
after integration over the collinear phase space regions.
These contributions are neglected by the above procedure.
However,
since these constant terms are not enhanced in any way
with respect to the leading logarithmic terms, we estimate the accuracy
of the approximation to be at the per cent level, which seems to be
good enough for most purposes.
Similarly, terms proportional to $m_e^2$ in the amplitude squared
of reaction (2) also yield small yet finite cross sections
for the RR polarization.
Indeed,
collinear electrons have some probability
to undergo a helicity flip \cite{Hagiwara}.
For unpolarized beams,
this effect is expected to be of the order
of the neglected terms proportional to $m_e^2/t^2$, discussed above.
As the photon in these collinear configurations
approaches its mass-shell, the corresponding contributions can
readily be estimated within the Weizs\"acker-Williams approximation
\cite{FS}.
Note that there are no singularities in the $\WW$ channel,
since the internal photon never couples to an on-shell fermion line.
In this case, the predictions made neglecting the electron mass
should be perfectly accurate. Also,
the potential collinear divergences
of $W^-$ or $Z^0$ radiation are automatically regulated by
the $W^-$ or $Z^0$ boson mass.

In order to obtain better numerical convergence of the Monte Carlo integration
we introduce the variable
\bea
\label{eq:variab}
   y = \half \ln {{1+2\Delta + \cos\theta}\over
                   {1+2\Delta - \cos\theta}}~,
\eea
where $\Delta$ is an appropriate regulator and $\theta$ is the angle
between the final state particle in question and the beam axis
\cite{Hagiwara}.
For the final electrons in reaction (2) and (3) we find
\bea
\label{eq:reg}
   \Delta = {m_e^2 \over s}
\eea
to be a good choice, while for the $Z^0$ bosons of reaction (3)
we take
\bea
\label{eq:regz}
   \Delta = {m_Z^2 \over s}~.
\eea
Using Eq.(\ref{eq:variab}), the collinear factor $(1 - \cos\theta )$ on the
right-hand-side of Eq.~(\ref{eq:t}) is replaced by
\bea
  1 - \cos\theta  = 2 {{1 + 2\Delta} \over
  {1+ \exp (- 2y)}} - 2 \Delta~.
\eea
The new variables $y$
and the remaining variables which are not discussed here
are generated according to flat distributions.
In this way we obtain stable results.

\section{Production Rates}

In this section, we present our results on polarized total cross
sections for the reactions (1) to (3).
The relevant input parameters used in our calculation
are given below. For the $Z^0$ and $W^-$ boson masses we take
$m_Z = 91.19~{\rm GeV}$ and $m_W = 80.3~{\rm GeV}$.
The corresponding value of the weak mixing angle
is given by $\sin^2\theta_W = 0.225$ .
Furthermore, for the fine structure constant we take its value
at the $Z^0$ pole, that is
$\alpha(m_Z^2) = 1/128.87$ \cite{Jegerlehner}.
This choice of scale is rather arbitrary,
as usual in $t$-channel processes.
Since one is dealing with $O(\alpha^4)$ processes the resulting
uncertainty is of the order of several per cent.
It is expected to exceed the errors due to the neglect
of the electron mass, pointed out in the previous section.
The value adopted for the Higgs boson mass is $m_H = 100~{\rm GeV}$.
Its exact value turns out to be irrelevant for reactions (1) and (2).
The same is true for reaction (3),
provided $m_H < 2m_Z$.
If $m_H > 2m_Z$ diagram 4 of Fig.~\ref{f3} develops a resonance
which is damped by the appropriate Higgs width.
We have not considered this possibility here.

The numerical predictions at a centre-of-mass energy of 500 GeV for polarized
and unpolarized beams are collected in Table~\ref{cs500}.
For a realistic integrated luminosity of 10 ${\rm fb^{-1}}$ and
unpolarized beams,
one can expect about 25 $\WW$, 100 $\WZ$ and 10 $\ZZ$ events.
{}From the leptonic and hadronic branching ratios
of the $W^-$, $Z^0$ and $\tau$,
%\begin{eqnarray*}
%\label{br}
%BR(W^- \to \ell^-\bar\nu_\ell) & \approx & 1/3 \\
%BR(W^- \to q\bar q') & \approx & 2/3 \\
%BR(Z^0 \to \ell^-\ell^+) & \approx & .1 \\
%BR(Z^0 \to \nu\bar\nu) & \approx & .2 \\
%BR(Z^0 \to q\bar q) & \approx & .7 ~,
%\end{eqnarray*}
one can easily estimate the total number of events with a given number
of neutrinos, charged leptons and hadronic jets.
In this context, though,
it is important to note
({\em cf.}\ next section)
that most of the final state leptons not coming from $W^-$ or $Z^0$
decays disappear along the beam-pipe,
whereas most of the decay products of the
(slow and rather isotropic)
gauge bosons should be observable.
For some important final states,
the number of events from reactions (1-3)
and some possible sources beyond the standard model
are listed in Table~\ref{en500},
assuming 10 fb$^{-1}$ of accumulated luminosity and unpolarized
beams. It is of course straightforward to perform
the same exercise with polarized beams.
As we shall show in the next section,
one can reduce the standard model background dramatically by
some simple kinematical cuts.

The energy dependence of the polarized cross sections is illustrated
in Fig.~\ref{tot}. Beyond $\sqrt{s} = 1$ TeV
the cross sections rise almost linearly.
Whereas for centre-of-mass energies above 1 TeV
the cross sections for reactions (1) and (2) become sizeable,
the cross sections of reaction (3) remain much smaller
over the whole energy range considered.
The striking differences in the relative magnitudes of the cross
sections apparent from Fig.~\ref{tot} result from the
differencies in boson-lepton couplings, the absence of triple
and quartic gauge boson couplings in (3) and the presence
of photon singularities in reactions (2) and (3). For example, we
have checked
that imposing an angular cut, which eliminates configurations in which
final state leptons are
collinear with the beam axis, diminishes the cross section of reaction
(2) to much larger extent than the cross section of reaction (1).
%The fact that the $LL$ beam polarization yields larger cross sections for
%reaction (2) than for reaction (1), in contrast to what one would naively
%expect from the relative magnitude of the $e^-e^-Z^0$ and
%$e^-\nu_{e}W^-$ couplings, can be traced back to the photon singularity
%discussed in the previous section.
%This is indeed expected from the weaker coupling of $Z^0$'s to electrons
%and the absence of triple gauge boson vertices.

To conclude this section,
we stress that important cancellations take place
between different gauge invariant subsets of Feynman diagrams.
At $\sqrt{s}=2$ TeV the total cross section can be three orders of
magnitude smaller than the individual contributions of some sets of
graphs. At higher energies,
the eventually ensuing numerical instabilities
might require the use of new calculational methods or approximations.

\section{Differential Distributions}

With the help of the relevant differential distributions,
we now characterize the typical topology
of the final states of reactions (1) to (3),
and examine which kinematical cuts can suppress them  most efficiently.
For definitness we concentrate on the process
$e_L^- e_L^- \rightarrow W^- Z^0 e^- \nu_e$
in the low ($\sqrt{s}=500$ GeV) and the high ($\sqrt{s}=2$ TeV)
energy regimes. The total cross sections are respectively
23 fb and 660 fb.
When a comparison is applicable, the leptons and gauge bosons
emerging from the other processes  have very similar behaviour.

An important feature of all reactions
is the strong tendency of the final leptons to emerge back-to-back
and close to the beam-pipe.
This is particularly true for the electrons in the processes
(2) and (3), because of the coupling to low-virtuality photons.
Moreover, with increasing collision energy this collinearity
becomes more and more extreme.

The angular distributions of the gauge bosons,
shown in Fig.~\ref{cw}, for the $W^-$, are also peaked along the beam
axis but by far not as strongly. At high energies,
the shape of the distribution remains essentially unchanged,
except for an enhancement in the small angle region.
The $Z^0$ distribution is only slightly more isotropic.
Obviously,
for the asymmetric combination of beam polarization $LR$,
the angular distributions are also asymmetric.
To be specific,
the final state electron and $W^-$ distributions
peak in the direction of the right-handed beam,
while the neutrino and $Z^0$ distributions
peak in the opposite directions.
However,
at high energies the $W^-$ and $Z^0$ distributions
become more symmetric.

In Fig.~\ref{cwz}, we plot the distributions
of the angles spanning the $W^-$ and the $Z^0$, and the $W^-$ and
the final electron.
Because the $Z^0$ couples stronger to the neutrino
than to the electrons,
it is mainly bremsstrahled off the outgoing neutrino line.
Therefore,
the $Z^0$ tends to be emitted parallel to the neutrino,
{\em i.e.} anti-parallel to the electron.
To conserve momentum,
the $W^-$ must then be emitted in the hemisphere of the electron.
While at high energies
this peaking becomes more pronounced,
one starts to observe at the same time
configurations with a small angle $W^-Z^0$-pair.
%This can be understood as the radiation of a $Z^0$
%off the neutrino in the process $e^-e^- \to e^-W^-\nu_e$,
%where a similar behaviour of the $W^-$ is observed \cite{wo1}.
The $e^-W^-$ correlation is very similar to the one predicted
for the reaction $e^-e^- \to e^-W^-\nu_e$ \cite{wo1}.

The energy distributions of the electron and the $W^-$
are displayed in Fig.~\ref{e}.
Again,
as in $e^-e^- \to e^-W^-\nu_e$,
irrespective of the centre-of-mass energy
the $W^-$ has a strong tendency to be emitted with moderate kinetic
energy. In contrast, the electron, and even more so the neutrino,
have high kinetic energies. In fact, at large centre-of-mass energies
the final leptons carry away most of the beam energy. At these
large energies, when the boson masses become negligible, the
energy distributions follow roughly the familiar Bremsstrahlung
spectra, {\em i.e.} $1/x$ for the gauge bosons and $1/(1-x)$
for the leptons.

Since hadronic decays of the gauge bosons
make up the bulk of the events ({\em c.f.}\ Table~\ref{en500}),
it may be interesting to describe some typical hadronic observables.
This is done in Fig.~\ref{wz},
where the total hadronic energy and transverse momentum distributions
are displayed side by side.
Independently of the centre-of-mass energy,
the transverse momentum remains peaked close to the value $p_{\perp}=
m_{W,Z}$.
As expected from the shapes of the angular distributions,
this confirms that the two gauge bosons
are preferentially emitted in opposite hemispheres.
The energy fraction carried by the massive gauge bosons
is of course substantial at low collision energy,
but shrinks noticeably as the centre-of-mass energy increases.
Nevertheless,
even at $\sqrt{s} = 500$ GeV
the tail of the energy distribution of the gauge boson pair
dies out rapidly in the upper energy range.
Consequently, imposing a lower limit on the total hadronic energy,
say $E_{\rm hadrons} > 0.8\sqrt{s}$,
would be very effective in reducing the event rate
to an unobservable level. On the other hand,
such a generous cut should not influence the number of hadronic events
expected from direct $W^-W^-$ production
via the $t$-channel exchange of a Majorana neutrino \cite{Minkowski},
even if one takes into account initial Bremsstrahlung.
Similarly,
requiring the transverse momentum not to exceed the experimental resolution
would fullfil the same purpose.

These observations about the angular and energy distributions
of the leptons and gauge bosons,
allow us to characterize roughly the event topologies
of processes (1) to (3) as follows:
the leptons carry away a major portion of the energy
along the beam directions,
while the gauge bosons are emitted with small velocity
and more isotropically. In the case of reaction (2),
the $W^-$ is emitted preferentially in the hemisphere of the electron,
while the $Z^0$ is to be found more often on the neutrino side.
%In essence,
%the two reactions (1) and (2) can be well described
%if seen as the radiation of an extra $W^-$ or $Z^0$
%``grafted'' to the process
%$e^-e^- \to e^-W^-\nu_e$.

\section{Conclusions}

We have performed a detailed analysis of three important mechanisms
for weak gauge boson pair production in $e^-e^-$ collisions,
within the framework of the standard model of electroweak interactions.
The total cross sections amount to only 1--10 fb for low collider energies,
but rise almost linearly with energy beyond $\sqrt{s} = 1$ TeV
and reach 200--700 fb at 2 TeV for $W^-W^-$ and $W^-Z^0$ production.
The event rates can then become substantial,
and could possibly serve as probes of quartic gauge couplings.
However,
these processes are unlikely to represent any danger
as backgrounds to more exotic reactions
which would signal departures from the standard model.
Indeed,
the final states are characteristic for radiative processes
and hence very distinct from exotic pair production
without accompanying fermion
or final states resulting from the decay of new heavy objects.

After this work was completed
we learned of other calculations
of the unpolarized total cross sections
for the processes considered here \cite{Barger,Boos}.
The predictions are in general agreement.

\bigskip
\bigskip
\bigskip

We are very grateful to Geert Jan van Oldenborgh for having provided us
with one of the phase space integration routines
used to evaluate the cross sections.

\vfill
\eject

\newpage
\bb{99}
\bi{Ginzburg} I.F.~Ginzburg, G.L.~Kotkin, V.I.~Telnov, Nucl.~Instr.
              Meth.~205 (1983) 47.
\bi{workshop} For general references see the Proceedings
	of the Munich, Annecy, Hamburg Workshops on
	$e^+e^-$ Collisions at 500 GeV (1991-93),
	DESY 92-123A, DESY 92-123B, DESY 93-123C.
\bi{Frampton} P.H.~Frampton, D.~Ng, \prd 45 (1992) 4240 (hep-ph9206244).
\bi{Minkowski} D.~London, G.~Belanger, J.N.~Ng, \plb 188 (1987) 155;\\
	C.A.~Heusch, P.~Minkowski, \np 416 (1994) 3.
\bi{KL} W.-Y.~Keung, L.~Littenberg, \prd 28 (1983) 1067.
\bi{COR1} F.~Cuypers, G.J.~van Oldenborgh, R.~R\"uckl, \np 409 (1993) 128
	(hep-ph9305287).
\bi{wo1} D.~Choudhury, F.~Cuypers, \plb 325 (1994) 500 (hep-ph9312308);\\
	D.~Choudhury, F.~Cuypers, Munich preprint MPI-Ph/94-24 (hep-ph9405212).
\bi{wo2} D.~Choudhury, F.~Cuypers, A.~Leike, Munich preprint MPI-Ph/94-23
(hep-ph9404362).
\bi{CKKR} F.~Cuypers, K.~Ko\l odziej, O.~Korakianitis, R.~R\"uckl,
         \plb 325 (1994) 243.
\bi{CKR} F.~Cuypers, K.~Ko\l odziej, R.~R\"uckl, to be published
         in the proceedings of the Zeuthen Workshop on Elementary
         Particle Theory, {\em Physics at LEP200 and beyond}, Teupitz,
         Germany, 10--15 April, 1994.
\bi{Kleiss} G.~Passarino, \np 237 (1984) 249;\\
            R.~Kleiss, W.J.~Stirling, \np 262 (1985) 235;\\
            D.-H.~Zhang, Z.~Xu, L.~Chang, \np 291 (1987) 392.
\bi{KZ} K.~Ko\l odziej, M.~Zra\l ek, \prd 43 (1991) 3619.
\bi{Hagiwara} K. Hagiwara {\em et al.}, \np 365 (1991) 544.
\bi{FS} B.~Falk, L.M.~Sehgal, \plb B325 (1994) 509.
\bi{Jegerlehner} H.~Burkhardt, F.~Jegerlehner, G.~Penso,
                 C.~Verzegnassi, \zp 43 (1989) 497.
\bi{Barger} V.~Barger, J.F.~Beacom, K.~Cheung, T.~Han,
	Madison preprint MAD-PH-779.
\bi{Boos} E.~Boos, M.~Duninin, private communication.
\eb

\vfill
\eject

\newpage

\begin{table}
\begin{center}
$\begin{array}{|c|c|c|c|}
\hline
     &    &    &    \\[-1.5mm]
 e^-e^- \rightarrow & \WW & \WZ & \ZZ \\
     &    &    &    \\[-1.5mm]
\hline
     &    &    &    \\[-1.5mm]
LL & 9.87    & 23.49    & 1.30    \\[1mm]
LR &   -     &  6.95    & 1.13    \\[1mm]
RR &   -     &    -     & 0.57   \\[1mm]
{\rm Unpolarized} & 2.47 & 9.35  & 1.03 \\[1.5mm]
\hline
\end{array}$
\caption{Total cross sections in femtobarns for various beam
polarizations at $\protect\sqrt{s} = 500$ GeV.}
\label{cs500}
\end{center}
\end{table}

\begin{table}
\begin{center}
\begin{tabular}{|c|c|c|}
\hline
observed & number of & exotic \\
final state & events & reactions \\
\hline
\hline
jets & 85 & $e^-e^- \to W^-W^-$ \cite{Minkowski} \\
\hline
charged leptons + jets & 34 & \\
$e^-$ + jets & 12 & $e^-e^- \to W^-W^-$ \cite{Minkowski} \\
 & & $e^-e^- \to e^-W^-\nu_e$ \cite{wo1} \\
\hline
charged leptons & 9 & \\
$e^-e^-$ & 1 & $e^-e^- \to \tilde e^-\tilde e^-$ \cite{KL,COR1} \\
 & & $e^-e^- \to e^-e^-$ \cite{wo2} \\
$\mu^-\mu^-$ & 1 & $e^-e^- \to X^{--}$ \cite{Frampton} \\
 & & $e^-e^- \to W^-W^-$ \cite{Minkowski} \\
 & & $e^-e^- \to \tilde\chi^-_1\tilde\chi^-_1$ \cite{COR1} \\
\hline
\end{tabular}
\caption{Number of background events expected from processes (1-3)
	to some specific reactions
	which might signal `new' physics.
	Only decay leptons are considered,
	while the leading leptons are likely to remain unobserved.
	We assume 10 fb$^{-1}$ integrated luminosity and unpolarized
        beams at $\protect\sqrt{s} = 500$ GeV.}
\label{en500}
\end{center}
\end{table}

\newpage

\end{document}